\RequirePackage{fix-cm}
\documentclass[twocolumn,epjc3]{svjour3}  
\smartqed  
\RequirePackage{graphicx}
\journalname{Eur. Phys. J. C}
\usepackage{cite}
\usepackage{amsmath}
\usepackage{amsfonts}
\usepackage{amssymb}
\usepackage{makeidx}
\usepackage{graphicx}
\usepackage{lmodern}
\usepackage{color}
\def\ZZ{{\Bbb Z}}
\begin{document}
\title{\textbf{{\color{blue} Perturbative solutions of the $f(R)$-theory of gravity 
in a central gravitational field and some applications}}}


\author{Nguyen Anh Ky \thanksref{e1,addr1,addr2}
        \and
        Pham Van Ky \thanksref{e2,addr3}
        \and 
        Nguyen Thi Hong Van \thanksref{e3,addr2,addr4}  
}
%
\thankstext{e1}{e-mail: anhky AT iop.vast.ac.vn}
\thankstext{e2}{e-mail: phamkyvatly AT gmail.com}
\thankstext{e3}{e-mail: nhvan AT iop.vast.ac.vn}

\institute{\small
\textit{Duy Tan university},\\
\small
\textit{K7/25 Quang Trung street, Hai Chau, Da Nang, Viet Nam.\\}\label{addr1}
          \and
          \small
\textit{Institute of physics},\\ 
\small
\textit{Vietnam academy of science and technology},\\ 
\small
\textit{10 Dao Tan, Ba Dinh, Hanoi, Viet Nam.\\}\label{addr2}
          \and
          \small
\textit{Graduate university of science and technology},\\ 
\small
{\it Vietnam academy of science and technology,}\\ 
\small
{\it 18 Hoang Quoc Viet, Cau Giay, Hanoi, Viet Nam.\\}\label{addr3}
\and 
\small
\textit{Institute for interdisciplinary research in science and education},\\ 
\small
\it{ICISE, Quy Nhon, Viet Nam.\\}\label{addr4}
}

\date{Received: date / Accepted: date}
%
%
\maketitle
\begin{abstract}

Exact solutions of an $ f(R) $-theory (of gravity) in a static central (gravitational) field 
have been studied in the literature quite well, but, to find and study exact solutions in the 
case of a non-static central field are not easy at all. There are, however, approximation methods 
of finding a solution in a central field which is not necessarily static. It is shown in this 
article that an approximate solution of an $f(R)$-theory in a general central field, which is not 
necessary to be static, can be found perturbatively around a solution of the Einstein equation 
in the general theory of relativity. In particular, vacuum solutions are found for $f(R)$ of 
general and some special forms. Further, applications to the investigation of a planetary motion 
and light's propagation in a central field are presented. An effect of an $f(R)$-gravity is also 
estimated for the SgrA*--S2 system. The latter gravitational system is much stronger than the 
Sun--Mercury system, thus the effect could be much stronger and, thus, much more measurable. 
\end{abstract}
\section{Introduction}

The General theory of Relativity (GR) \cite{1, 2} announced in 1915 was theoretically developed 
by A. Einstein, and it has been experimentally verified as an excellent theory of gravitation.  
In particular, the GR was once again confirmed triumphantly by recent detections of 
gravitational waves (see, for example, \cite{Abbott:2016blz,TheLIGOScientific:2017qsa}).  
The GR is governed by the Einstein equation \cite{1, 2, P1}
\begin{align*}
R_{\mu\nu}-\frac{1}{2}Rg_{\mu\nu}=-\frac{8\pi{G}}{c^4}T_{\mu\nu},
\end{align*}
obtained from the Lagrangian  $ {\cal L}_G= R  $. This equation can describe very well gravitational 
phenomena of the normal matter, but it  ineffectively describes other phenomena such as the Universe's 
accelerated expansion (supposed to be explained by the introduction of the concept of the so-called 
dark energy or cosmological constant), dark matter, cosmic inflation, quantum gravity, etc. One of the 
simplest suggestions for solving the dark energy problem is adding the cosmological constant $ \Lambda $ 
to the Lagrangian, that is,  
${\cal L}_G=R-2\Lambda $, leading to the equation of motion  \cite{1, P1}
\begin{align*}
R_{\mu\nu}-\frac{1}{2}Rg_{\mu\nu}+\Lambda g_{\mu\nu}=-\frac{8\pi{G}}{c^4}T_{\mu\nu}.
\end{align*} 
According to the latter equation, the Universe would acceleratedly expand. However, there is also 
room for doubt in this case  (see \cite{7,DeFelice:2010aj,thomas} for more discussions).  \\

A more general theory\footnote{There are also other models extending the GR, however, they 
are not in the scope of the present paper (see \cite{DeFelice:2010aj,thomas} and references therein, 
for listing some of them).}, which can be used to solve the above-mentioned  problems and explain 
some other phenomena 
in cosmology is that with Lagrangian  ${\cal L}_G= f(R)$, where $ f(R) $ is a scalar function of the scalar curvature 
$ R $. The equation of motion now becomes  \cite{DeFelice:2010aj, thomas, Capozziello:2011et}
 \begin{align*}
f'(R)R_{\mu\nu}-g_{\mu\nu}\square f'(R)+\nabla_{\mu}\nabla_{\nu} f'(R)-\frac{1}{2}f(R)g_{\mu\nu}\\=-kT_{\mu\nu},
\end{align*}
where $k= \frac{8\pi G}{c^4}$, $ \square = \nabla_{\mu}\nabla^{\mu} $ and $ \nabla_{\mu} $ is the covariant derivative. 
This theory is called $f(R)$-theory of gravity or just $f(R)$-gravity or $f(R)$-theory for short. Nowadays, this theory 
is becoming a hot topical issue and attracting much interest of many cosmologists (for review, see, for example, 
\cite{DeFelice:2010aj, thomas, Capozziello:2011et, Amendola:2006kh, Wei:2015xax, Nojiri:2003ft}). There are a lot of 
variants of the $f(R)$-theory such as those with $ f(R)=R+\lambda R^2 $ or $ f(R)=R-\frac{\lambda}{R^n} $, etc., 
each them can explain some of the cosmological phenomena but none of them is perfect \cite{DeFelice:2010aj} -- 
\cite{Muller:2014qja}. In general, to find a solution, especially, an exact one, of an $f(R)$-theory is very difficult, 
even impossible. To simplify the situation, some reasonable conditions are sometimes imposed so that 
approximate solutions can be found. One of such conditions could be that of a spherical symmetry which is a quite 
good approximation in many cases. Here, a general $f(R)$-theory will be considered in a spherically symmetric 
(gravitational) field called usually a central field. \\

Exact solutions of the $ f(R) $-theory in a static central field are studied in \cite{Multamaki:2006zb, Kainulainen:2007bt, 
Shojai:2011yq, Sharif:2011uf, Sebastiani:2010kv, Erickcek:2006vf} but there are also approximation methods for central 
fields which are not necessarily static \cite{Arbuzova:2013pta, Stabile:2010zk, Capozziello:2009vr}. In this article, 
approximate solutions of the $f(R)$-theory for a general and some special cases in a general central field are 
found by perturbation around the Einstein equation. 
Then, we can use the obtained solutions to calibrate parameters of orbits of planets.\\

In this article the following conventions are used:

\begin{itemize}
	
\item Metric signature in Minkowski space: ($ +,  -,  -,  - $), that is, 
the infinitesimal distance is calculated as 
\begin{align*}
ds^2&=\eta_{\mu\nu}dx^{\mu}dx^{\nu}= dx^0dx_0+dx^idx_i\\
&=c^2dt^2-dx^2-dy^2-dz^2, 
\end{align*} 
with Latin letters used for three-dimensional 
spatial indices, and, Greek letters used for four-dimensional space-time indices.
\vspace*{2mm}

\item  Riemann curvature tensor: 
\begin{eqnarray*}
R^{\alpha}_{~\mu\beta\nu}=\frac{\partial \Gamma^\alpha_{\mu\beta}}{\partial x^\nu} - 
\frac{\partial \Gamma^\alpha_{\mu\nu}}{\partial x^\beta} + 
\Gamma^\alpha_{\sigma\nu}\Gamma^\sigma_{\mu\beta} -  
\Gamma^\alpha_{\sigma\beta}\Gamma^\sigma_{\mu\nu}.
\end{eqnarray*}
\vspace*{1mm}

\item Rank-2 curvature tensor (Ricci tensor):\\
$R_{\mu\nu}=R^\alpha_{~\mu\alpha\nu}.$
\vspace*{2mm}

\item Scalar curvature:  $ R=g^{\mu\nu}R_{\mu\nu} $.
\vspace*{2mm}

\item Energy-momentum tensor of a macroscopic object: 
\begin{eqnarray*}
T_{\mu\nu}
=\frac{1}{c^2}(\varepsilon + p)u_\mu u_\nu - pg_{\mu\nu}, 
\end{eqnarray*}
where $ u^\mu 
= \frac{dx^\mu}{d\tau}=c\frac{dx^\mu}{ds} $, while $ \varepsilon$ and $p$ 
are the energy density and the pressure, respectively.
\end{itemize}	

In the next section we will consider a general $f(R)$-theory in a central field and its perturbative solutions. 
In particular, solutions in vacuum are also investigated for a general form and some special forms of $f(R)$. 
Sec. 3 is devoted to applications of the obtained solutions to investigating a planet's and light's motion in 
a central field. Some comments and conclusion will be made in Sect. 4. Finally, a proof of formula \eqref{36} 
is exposed in the appendix.

\section{$f(R)$-theory and perturbative solutions}

Now we consider a system of matter in a gravitational field. If the gravitational field's Lagrangian is 
${\cal L}_G = R$ and the matter Lagrangian is ${\cal L}_M$, the system's action has the form 
\begin{align}
S&=S_G + S_M\nonumber\\
&=\frac{c^3}{16\pi{G}}\int{R{\sqrt{-g}} {d^4x}}+\frac{1}{c}\int{{\cal L}_M{\sqrt{-g}}{d^4x}}. 
\label{1}
\end{align}
The Einstein equation obtained from this action \cite{1},\cite{2} is   
\begin{align} 
R_{\mu\nu}-\frac{1}{2}g_{\mu\nu}R=-\frac{8\pi{G}}{c^4}T_{\mu\nu},\label{2}
\end{align}
where $T_{\mu\nu}$ is the energy-momentum tensor of matter  
\begin{align}
 T_{\mu\nu}:=&\frac{+2}{\sqrt{-g}}\frac{\delta S_M}{\delta g^{\mu\nu}}\\ 
 =&-{\cal L}_Mg_{\mu\nu}
 +2\frac{\partial{{\cal L}_M}}{\partial{g^{\mu\nu}}}-\frac{2}{\sqrt{-g}}
 \frac{\partial}{\partial{x^{\alpha}}}\left( \frac{\partial{({\cal L}_M\sqrt{-g})}}
 {\partial{\frac{\partial{g^{\mu\nu}}}{\partial{x^{\alpha}}}}}\right).\nonumber
\end{align}  
Taking a trace of \eqref{2}, we get   
\begin{align}
R=\frac{8\pi{G}}{c^4}T \label{R}
\end{align}
with $T=T^\mu_\mu$, and the equation \eqref{2} becomes 
\begin{align} 
R_{\mu\nu}=-\frac{8\pi{G}}{c^4}\left (T_{\mu\nu}-{1\over 2} g_{\mu\nu}T\right). \label{RT}
\end{align} 
For ${\cal L}_G = f(R)$, the system's action is 
\begin{align*}
S &= S_G + S_M\\
  &=\frac{c^3}{16\pi{G}}\int{f(R){\sqrt{-g}} {d^4x}}+\frac{1}{c}\int{{\cal L}_M{\sqrt{-g}}{d^4x}} 
\end{align*}
leading to the equation of motion \cite{DeFelice:2010aj, thomas, Capozziello:2011et} 
\begin{align}
f'(R)R_{\mu\nu}-g_{\mu\nu}\square f'(R)+\nabla_{\mu}\nabla_{\nu} f'(R)-\frac{1}{2}f(R)g_{\mu\nu}\nonumber\\
=-kT_{\mu\nu}. 
\label{6}
\end{align}
If the $f(R)$-theory deffers from the Einstein theory (when $ f(R)=R $) just slightly, we can write $f(R)$ 
in the form 
\begin{align}
f(R)=R+\lambda h(R) \label{5},
\end{align}
where $ h(R) $ is a scalar function of  $ R $ and $\lambda$ is a parameter such that $ \lambda h(R)$ 
and its derivatives compared with $R$ are a very small. 
Substituting  
\eqref{5} into \eqref{6} we obtain 
\begin{align*}
[1+\lambda h'(R)]R_{\mu\nu}-g_{\mu\nu}\square [\lambda h'(R)]+\nabla_{\mu}\nabla_{\nu}[\lambda h'(R)]\\
-\frac{1}{2}g_{\mu\nu}[R+\lambda h(R)]=-kT_{\mu\nu},
\end{align*}
or
\begin{align}
R^{\mu}_{~\nu}-\frac{1}{2}\delta^{\mu}_{~\nu}R+\lambda h'(R)R^{\mu}_{~\nu}-\frac{\lambda}{2}\delta^{\mu}_{~\nu}
h(R)-\lambda \delta^{\mu}_{~\nu}\square h'(R) \nonumber \\ 
+\lambda \nabla^{\mu}\nabla_{\nu}h'(R)=-kT^{\mu}_{~\nu}. \label{8}
\end{align}
We solve the latter equation by a perturbation method basing on the fact that this equation differs from the 
Einstein equation by small perturbative terms (the last four terms on the left-hand side of the last equation). 
Substituting a solution of the Einstein equation, i.e.,  \eqref{R} and \eqref{RT}, 
$$R=kT, ~~  R^{\mu}_{~\nu}=-k(T^{\mu}_{~\nu}-\frac{1}{2}\delta^{\mu}_{~\nu}T),$$
into the perturbative terms in Eq. \eqref{8}, 
\begin{align}
R^{\mu}_{~\nu}-\frac{1}{2}\delta^{\mu}_{~\nu}R-\lambda k h'(kT)(T^{\mu}_{~\nu}
-\frac{1}{2}\delta^{\mu}_{~\nu}T) -\frac{\lambda}{2}\delta^{\mu}_{~\nu}h(kT)
\nonumber \\
-\lambda \delta^{\mu}_{~\nu}\square^E h'(kT)+\lambda \nabla^{\mu}\nabla_{\nu}^Eh'(kT)
=-kT^{\mu}_{~\nu}, \label{9}
\end{align}
we solve the latter perturbatively at the first order, where $h'(kT)=\frac{\partial h(kT)}{\partial (kT)}$ 
and the superscript $E$ in the covariant differentiations means that the metric tensor  $g_{\mu\nu}$ is 
taken in the Einstein equation solutions. If we solve the above equation in vacuum ($T^{\mu}_{~\nu}=0$, 
$T=0$), then $h(kT)$ and $h'(kT)$ are constants, hence their differentiations are equal to zero, the 
equation \eqref{9} becomes
\begin{align}
R^{\mu}_{~\nu}-\frac{1}{2}\delta^{\mu}_{~\nu}R=\frac{\lambda}{2}\delta^{\mu}_{~\nu}h(kT)
=\frac{\lambda}{2}\delta^{\mu}_{~\nu}h(0). \label{10}
\end{align}
Note: If $h(0)=0$ the perturbative equation \eqref{10} is similar to the Einstein equation in vacuum. 
But, there is a fundamental difference. With the Einstein equation in a central field, a solution in 
vacuum is stationary and determined upto a constant (of time) even when the central field is not stationary. 
In a spherically symmetric $f(R)$-theory, however, as seen later, a solution in general is not stationary.  
Furthermore, the integration constant in the solution of the Einstein equation can be found by taking a 
limit at the classical gravitational potential $\varphi =  -\frac{GM}{r}$, but a similar step cannot be done 
with the  $f(R)$-theory as the classical gravitational potential may be different from $\varphi = -\frac{GM}{r}$, 
though little. Hence, we will solve Eq. \eqref{9} in a general way, not only in vacuum.\\

We are using a spherically symmetric metric in the shape of the Schwarzschild metric \cite{2}, 
\begin{align}
ds^2=e^{u(r,t)}{dx^o}^2-e^{v(r,t)}dr^2-r^2(d\theta^2+\sin^2\theta d\varphi^2), \label{11}
\end{align}
with the following non-zero metric elements 
\begin{align*}
g_{00}=e^{u(r,t)}, ~g_{11}= -e^{v(r,t)},\\
~g_{22}= -r^2 ,~ g_{33}=-r^2\sin^2\theta
\end{align*}
(writing $ g_{00}=e^{u(r,t)} $ does not mean it always positive because $u(r,t)$ can be complex). 
With the given metric we can calculate any element  
of the Ricci tensor \cite{2}, say $R^{1}_{~0}$,
\begin{align}
R^{1}_{~0}= \frac{e^{-v(r,t)}}{r}\frac{\partial{v(r,t)}}{\partial ct}, \label{12}
\end{align}
which, when inserted in Eq. \eqref{6} for vacuum, gives
\begin{align}
f'(R)\frac{e^{-v(r,t)}}{r}\frac{\partial{v(r,t)}}{\partial ct}
=-\nabla^1\nabla_0f'(R). \label{13}
\end{align}
Considering the case of a central gravitational field, we see that the 
Einstein theory in vacuum, as well known, 
is always stationary, but,   
a general $f(R)$-theory, with $ f(R)\neq R $, is not stationary in vacuum. 
It can be seen from the fact that the right-hand side of \eqref{13} in 
general is not zero, therefore,  $\frac{\partial{v(r,t)}}{\partial t}\neq0$, 
thus, the metric may varies with time. However, the non-stationarity does 
not show up at the first order of perturbation by using $R^{1}_{~0}$ in the 
approximate equation \eqref{9}. To see the non-stationarity appearing at 
the first order of perturbation, it is enough to use $ R^0_{~0} $ and 
$ R^1_{~1}$ in the latter equation.\\

With the metric \eqref{11} we get \cite{2} 
\begin{align}
R^0_{~0}-\frac{1}{2}R=e^{-v(r,t)}\left[ \frac{1}{r^2}-\frac{v'(r,t)}{r}\right] -\frac{1}{r^2} ~, \label{14}
\end{align}
where $ v'(r,t)=\frac{\partial v(r,t)}{\partial r} $. Comparing  \eqref{14} with \eqref{9}  we obtain
\begin{align*}
e^{-v(r,t)}\left[ \frac{1}{r^2}-\frac{v'(r,t)}{r}\right] -\frac{1}{r^2}=-kT^0_{~0}+\frac{\lambda}{2}h(kT)\nonumber \\
+\lambda \nabla^i \nabla_i^E h'(kT)+\lambda k \left(T^0_{~0}-\frac{T}{2}\right)h'(kT),
\end{align*}
where $ \nabla^i \nabla_i = \square -\nabla^0 \nabla_0 $,
or
\begin{align}
e^{-v(r,t)}\frac{v'(r,t)}{r}+\frac{1}{r^2}\left[1-e^{-v(r,t)}\right]=kT^0_{~0}-\frac{\lambda}{2}h(kT)\nonumber \\
-\lambda \nabla^i \nabla_i^E h'(kT)-\lambda k\left (T^0_{~0}-\frac{T}{2}\right )h'(kT). \label{15}
\end{align}
If we write $v(r,t)$ in the form 
\begin{align}
v(r,t)=-\mbox{ln}\left[ 1+\frac{c(r,t)}{r}\right] , \label{16}
\end{align}
we get from \eqref{15}
\begin{align*}
-\frac{c'(r,t)}{r^2}=kT^0_{~0}-\frac{\lambda}{2}h(kT)-\lambda \nabla^i \nabla_i^E h'(kT)\nonumber \\
-\lambda k(T^0_{~0}-\frac{T}{2})h'(kT).
\end{align*}
Integrating the latter equation 
\begin{align}
c(r,t)=-\int^r_0\left[ kT^0_{~0}-\frac{\lambda}{2}h(kT)
-\lambda \nabla^i \nabla_i^E h'(kT)\right. \nonumber \\
\left. -\lambda k(T^0_{~0}-\frac{T}{2})h'(kT)\right] r'^2dr', \label{17}
\end{align}
and substituting the result \eqref{17} into \eqref{16}, we obtain
\begin{align}
v(r,t)=-\mbox{ln}\left\lbrace 1-\frac{1}{r}\int_0^r\left[ kT^0_{~0}
-\frac{\lambda}{2}h(kT)\right.\right.\nonumber \\
\left.\left. -\lambda \nabla^i \nabla_i^E h'(kT)-\lambda k(T^0_{~0}
-\frac{T}{2})h'(kT)\right] r'^2dr'\right\rbrace, \label{18}
\end{align}
where $ T^0_{~0}=T^0_{~0}(r',t) $ and  $ T=T(r',t) $. \\

Let us now calculate the integrand $\nabla^i \nabla_i^E h'(kT)$ 
in \eqref{18}. Because of the spherical symmetry, $T$, thus, $ h'(kT) $ does 
not depend on $ \theta$ and $\varphi $, but $r$ and $t$ only. Putting only non-vanishing 
elements of $ g_{\mu\nu}$ and $ \Gamma^\alpha_{\mu\nu} $ in the intergrand,  we have
 \begin{align}
 \nabla^i\nabla_ih'(kT)=g^{11}\partial_1\partial_1h'(kT)-g^{11}
 \Gamma^0_{11}\partial_oh'(kT)\nonumber \\ 
 -g^{ij}\Gamma^1_{ij}\partial_1h'(kT), \label{19}
 \end{align}
 here $ \partial_0=\frac{\partial}{\partial x^0}=\frac{\partial}{\partial ct} $, 
 $ \partial_1=\frac{\partial}{\partial r} $ and  $ g^{ij}\Gamma^1_{ij}
 =g^{11}\Gamma^1_{11}+g^{22}\Gamma^1_{22}+g^{33}\Gamma^1_{33} $. 
 On the other hand, also because of the spherical symmetry, we have
 \begin{align}
 g^{11}\Gamma^0_{11}=-\frac{g^{11}g^{00}}{2}\frac{\partial g_{11}}{\partial x^0}
 =-\frac{g^{11}g^{00}}{2}\frac{\partial \frac{1}{g^{11}}}{\partial x^0}
 =\frac{1}{2c}\frac{g^{00}}{g^{11}}\frac{\partial g^{11}}{\partial t}, \label{20}
 \end{align}
 \begin{align}
 g^{ij}\Gamma^1_{ij}= ~ & g^{11}g^{11}\frac{\partial g_{11}}{\partial x^1}
 -\frac{1}{2}g^{11}g^{ii}\frac{\partial g_{ii}}{\partial x^1}\nonumber \\
 = ~ & g^{11}g^{11}\frac{\partial \frac{1}{g^{11}}}{\partial x^1}
 -\frac{1}{2}g^{11}g^{ii}\frac{\partial g_{ii}}{\partial x^1}\nonumber \\
 = & -\frac{1}{2}\frac{\partial g^{11}}{\partial r}-\frac{2}{r}g^{11}. \label{21}
 \end{align}
 Finally, substitutions of \eqref{20} and \eqref{21} into \eqref{19} give  
 \begin{align}
  \nabla^i\nabla_i^Eh'(kT)=-\frac{1}{2c^2}\frac{g^{00}_E}{g^{11}_E}\frac{\partial g^{11}_E}
  {\partial t}\frac{\partial h'(kT)}{\partial t}+g^{11}_E\frac{\partial^2h'(kT)}{\partial r^2}\nonumber \\ 
+\left( \frac{2}{r}g^{11}_E+\frac{1}{2}\frac{\partial g^{11}_E}{\partial r}\right) 
\frac{\partial h'(kT)}{\partial r}. \label{22}
 \end{align}
Here, as said before, the subscript $E$ indicates the Einstein limit.
\textbf{\subsection{Vacuum solutions}}
Now we consider solutions in vacuum for a body-gravitation source of radius $R_0$ which in general depends on time, $R_0 = R_0(t)$. 
Because of considering solutions in vacuum, we can neglect the  pressure.
It means that the tensor $T$ has $T^0_{~0}$ as the only non-zero component, and Eq. \eqref{18} becomes 
\begin{align}
v(r,t)=&-\mbox{ln}\left\lbrace 1-\frac{1}{r}\int_0^r\left[ kT^0_{~0}-
\frac{\lambda}{2}h(kT^0_{~0})
\right.\right. \nonumber \\
&\left.\left.-\frac{\lambda}{2}kT^0_{~0}h'(kT^0_{~0})
-\lambda\nabla^i\nabla_i^Eh'(kT^0_{~0})\right] r'^2dr'\right\rbrace, \label{23}
\end{align}
with  $ \nabla^i\nabla_i^Eh'(kT^0_{~0}) $ calculated by \eqref{22}. As $ T^0_{~0}=0 $ 
in vacuum, the first integration in \eqref{23} spreads only between 0 and $R_0(t)$ and gives 

\begin{align}
\int^{R_0(t)}_0 kT^0_{~0}(r',t)~r'^2dr'= \frac{kMc^2}{4\pi},
 \label{24a}
\end{align}
with $M$ being the mass of the body-gravitation source. 
Assuming that the body-gravitation source is uniform we have 
\begin{align} 
M= {\frac{4}{3}}\pi {[R_0(t)]}^3\times \rho, 
\label{24b}
\end{align}
where, $ \rho $ is the mass density which is independent from coordinates, and, thus, 
\begin{align} 
T^0_{~0}= {M c^2\over {\frac{4}{3}}\pi {[R_0(t)]}^3}.
 \label{24}
\end{align}
Putting  \eqref{24a} in \eqref{23} we get 
\begin{align}
v(r,t)=-\mbox{ln}\left\lbrace 1-\frac{kc^2M}{4\pi r}+\frac{\lambda}{r}\int^r_0\left[\frac{1}{2}h(kT^0_{~0})
\right.\right. \nonumber \\
\left.\left.+\frac{1}{2}kT^0_{~0}h'(kT^0_{~0})
 +\nabla^i \nabla_i^E h'(kT^0_{~0})\right] r'^2dr'\right\rbrace. \label{26}
\end{align}
Next we calculate $ u(r, t) $ in $ g_{00}(r, t $). Considering \eqref{9} in vacuum, see \eqref{10},
\begin{align}
R^{0}_{~0}-\frac{1}{2}R=R^{1}_{~1}-\frac{1}{2}R, \label{27}
\end{align}
and using the metric \eqref{11}, we have  \cite{2} 
\begin{align}
R^0_{~0}-\frac{1}{2}R=-e^{-v(r,t)}\left[ \frac{v'(r,t)}{r}-\frac{1}{r^2}\right] -\frac{1}{r^2},  \label{28}\\
R^1_{~1}-\frac{1}{2}R=e^{-v(r,t)}\left[ \frac{u'(r,t)}{r}+\frac{1}{r^2}\right] -\frac{1}{r^2},  \label{29}
\end{align}
and, thus, $ u(r,t)=-v(r,t)$. Therefore, in vacuum, $u(r,t)$ takes the form 
\begin{align}
u(r,t)=&\mbox{ln}\left\lbrace 1-\frac{kc^2M}{4\pi r}
+\frac{\lambda}{r}\int^r_0\left[\frac{1}{2}h(kT^0_{~0})\right.\right. \nonumber \\ 
&\left.\left.+\frac{1}{2}kT^0_{~0}h'(kT^0_{~0}) +\nabla^i \nabla_i^E h'(kT^0_{~0})\right] r'^2{dr'}\right\rbrace.  \label{30}
\end{align}
In conclusion,  starting from $ {\cal L}_G=f(R)=R+\lambda h(R) $ and the Schwarzschild metric 
[thus \eqref{11}, \eqref{26} and \eqref{30}], we obtain a perturbative solution in vacuum 
\begin{align}
g_{00}(r,t)=&1-\frac{kc^2M}{4\pi r}+\frac{\lambda}{r}\int^r_0\left[\frac{1}{2}h(kT^0_{~0})
\right. \nonumber \\ 
&\left. +\frac{1}{2}kT^0_{~0}h'(kT^0_{~0})+\nabla^i \nabla_i^E h'(kT^0_{~0})\right] r'^2dr',     \label{31}
\end{align}
\begin{align}
g_{11}(r,t)=-\left\lbrace 1-\frac{kc^2M}{4\pi r}+\frac{\lambda}{r}\int^r_0\left[\frac{1}{2}h(kT^0_{~0})\right.\right. \nonumber \\
\left.\left. +\frac{1}{2}kT^0_{~0}h'(kT^0_{~0})+\nabla^i \nabla_i^E h'(kT^0_{~0})\right] r'^2dr'\right\rbrace ^{-1}, \label{32}
\end{align}
\begin{align}
g_{22}=-r^2, \label{33}
\end{align}
\begin{align}
g_{33}=-r^2sin^2\theta, \label{34}
\end{align}
with $ k=\frac{8\pi G}{c^4} $, $ T^0_{~0}=T^0_{~0}(r',t) $, $ h'(kT^0_{~0})
=\frac{\partial h(kT^0_{~0})}{\partial (kT^0_{~0})} $ and 
$ \nabla^i \nabla_i^E h'(kT^0_{~0}) $ 
calculated in \eqref{22}.
Far away from the body-gravitation source $ T^0_{~0} $ can be considered depending on the time $ t $ 
only (at a long distance, the density of the body-gravitation source can be considered homogeneous), 
that means the last two terms of \eqref{22} vanishing (see more details in the appendix), 
\begin{align}
\nabla^i \nabla_i^E h'(kT^0_{~0})=-\frac{1}{2c^2}\frac{g^{00}_E}{g^{11}_E}
\frac{\partial g^{11}_E}{\partial t}\frac{\partial h'(kT^0_{~0})}{\partial t}. \label{35}
\end{align}
 \begin{eqnarray}
&&\int_0^{R_0(t)}\nabla^i \nabla_i^E h'(kT^0_{~0})r'^2dr'\nonumber \\
&&~~~~~~~~~~~~~~~~~~~\approx h''(kT^0_{~0})\left[ \frac{\partial}{\partial t}
\frac{M}{[R_0(t)]^3}\right]^2 \alpha (t), \label{36}
\end{eqnarray}
where, 
\begin{align}
\alpha (t)=&\frac{3k^2c^2R_0(t)}{256\pi^2[\xi (t)]^4}\left\lbrace \frac{3}{\xi(t)R_0(t)}
\arcsin[\xi (t) R_0(t)]\right.\nonumber \\
 &\left. -\left( 3+2[\xi(t)R_0(t)]^2\right) \sqrt{1-[\xi(t)R_0(t)]^2}\right\rbrace
 \nonumber \\
&\times \left( 1-[\xi (t)R_0(t)]^2\right)^{-3/2}, \label{37}
\end{align}
with  
\begin{align}
\xi^2 (t)=\frac{kMc^2}{4\pi [R_0(t)]^3}. \label{37a}
\end{align}
Substituting  \eqref{36} into  \eqref{31} -- \eqref{34} we find a solution at a 
distant point from the body-gravitation source: 
\begin{align}
g_{00}(r,t)=&1-\frac{kc^2M}{4\pi r}\nonumber \\ 
&+\frac{\lambda}{2r}\int^r_0\left[h(kT^0_{~0})+kT^0_{~0}h'(kT^0_{~0})\right] r'^2dr'\nonumber \\
&+\frac{\lambda h''(kT^0_{~0})}{r}\left[ \frac{\partial}{\partial t}
\frac{M}{[R_0(t)]^3}\right]^2 ~\alpha (t), \label{38}
\end{align}
\begin{align}
g_{11}(r,t)=&-\left\{1-\frac{kc^2M}{4\pi r}\right.\nonumber \\
&+\frac{\lambda}{2r}\int^r_0\left[h(kT^0_{~0})+kT^0_{~0}h'(kT^0_{~0})\right] r'^2dr'\nonumber \\
&\left.+\frac{\lambda h''(kT^0_{~0})}{r}\left[ \frac{\partial}{\partial t}\frac{M}{[R_0(t)]^3}\right]^2
~\alpha (t)\right\}^{-1}, \label{39}
\end{align}
\begin{align}
g_{22}=-r^2, \label{40}
\end{align}
\begin{align}
g_{33}=-r^2\sin^2\theta, \label{41}
\end{align}
here \eqref{24} is used for both inside and outside the integral, and 
$ h''(kT^0_{~0})=\frac{\partial^2h(kT^0_{~0})}{\partial(kT^0_{~0})^2} $.  
Note that though $ T^0_{~0} $  depends on time $ t $ only, one should be 
careful when bring $ h(kT^0_{~0})+kT^0_{~0}h'(kT^0_{~0}) $ out of the integral. 
If $ h(kT^0_{~0})+kT^0_{~0}h'(kT^0_{~0})=0 $ in vacuum, the integral is performed 
inside the body-gravitation source, but there are also cases when 
$ h(kT^0_{~0})+kT^0_{~0}h'(kT^0_{~0}) $ 
is not zero in vacuum (see below). Now, in the next subsection, applying the 
latest formulas, we consider some special cases.\\
\textbf{\subsubsection{The case $ f(R)=R-2\lambda$ (model I)}}
 In this case we have $ h(R)=-2 $ leading to $ h(kT^0_{~0})=-2 $, $ h'(kT^0_{~0})=0 $ 
 and the formulas from \eqref{38} to \eqref{41}  can be calculated easily as
\begin{align}
g_{00}(r,t)=1-\frac{kc^2M}{4\pi r}-\frac{\lambda r^2}{3}, \label{42}
\end{align}
\begin{align}
g_{11}(r,t)=\frac{-1}{1-\frac{kc^2M}{4\pi r}-\frac{\lambda r^2}{3}}, \label{43}
\end{align}
\begin{align}
g_{22}=-r^2, \label{44}
\end{align}
\begin{align}
g_{33}=-r^2sin^2\theta. \label{45}
\end{align}
It is exactly the solution of the modified Einstein equation with a cosmological constant $\lambda$.\\
\textbf{\subsubsection{The case $ f(R)=R+\lambda R^b $,  $ b>0 $ (model II)}}
Thus, $ h(R)=R^b $ and  $h'(R)= bR^{b-1} $, $ h''(R)=b(b-1)R^{b-2} $, the 
formulas  \eqref{38} --  \eqref{41} become
\begin{align}
g_{00}=&1-\frac{kc^2M}{4\pi r}+\frac{\lambda (b+1)k^b }{2r}\int^{R_o(t)}_0{[T^0_{~0}]}^br'^2dr'
\nonumber \\
&+\frac{\lambda}{r}b(b-1)k^{b-2}(T^0_{~0})^{b-2}\left[ \frac{\partial}
{\partial t}\frac{M}{[R_o(t)]^3}\right]^2 ~\alpha (t), \label{46}
\end{align}
\begin{align}
g_{11}=&-\left\{1-\frac{kc^2M}{4\pi r}+\frac{\lambda (b+1)k^b }{2r}
\int^{R_o(t)}_0{[T^0_{~0}]}^br'^2dr'\right.\nonumber \\ 
& \left.+ \frac{\lambda}{r} b(b-1)k^{b-2}(T^0_{~0})^{b-2}\left[ 
\frac{\partial}{\partial t}\frac{M}{[R_o(t)]^3}\right]^2 
~\alpha (t)\right\}^{-1}, \label{47}
\end{align}
\begin{align}
g_{22}=-r^2, \label{48}
\end{align}
\begin{align}
g_{33}=-r^2sin^2\theta. \label{49}
\end{align}
Further, applying \eqref{24} we have
\begin{align}
&g_{00}(r,t)=1-\frac{kc^2[M-\lambda M_1(t)-\lambda M_2(t)]}{4\pi r}, \label{50}\\
&g_{11}(r,t)=\frac{-1}{1-\frac{kc^2[M-\lambda M_1(t)-\lambda M_2(t)]}{4\pi r}}, \label{51}\\
&g_{22}=-r^2, \label{52}\\
&g_{33}=-r^2\mbox{sin}^2\theta. \label{53}
\end{align}
Here
\begin{align}
&M_1(t)=\frac{4\pi}{kc^2}\frac{(b+1)c^{2b}(kM)^b }{3^{1-b}~2^{2b+1}~{\pi}^b[R_o(t)]^{3b-3}}, \label{53a}\\
&M_2(t)=\frac{4\pi}{kc^2}\frac{b(b-1)c^{2b-4}(3kM)^{b-2}\left[ 
\frac{\partial}{\partial t}\frac{M}{[R_o(t)]^3}\right]^2 \alpha (t)}{(4\pi)^{b-2}[R_o(t)]^{3b-6}}. \label{53b}
\end{align}
\subsubsection{\textbf{The case $f(R)= R^{1+\varepsilon}$ (model III)}}
Here $\varepsilon$ is an infinitesimally small number. In this case   
$ \lambda h(R) = R^{1+\varepsilon}-R $ and  $\lambda h'(R) = (1+\varepsilon )
R^\varepsilon-1$, $ \lambda h''(R)=\varepsilon (\varepsilon+1)R^{\varepsilon-1} $. 
Similarly,  we obtain the corresponding metric tensor
\begin{align}
&g_{00}(r,t)=1-\frac{kc^2[M-\lambda M_1(t)-\lambda M_2(t)]}{4\pi r}, \label{54}\\
&g_{11}(r,t)=\frac{-1}{1-\frac{kc^2[M-\lambda M_1(t)-\lambda M_2(t)]}{4\pi r}}, \label{55}\\
&g_{22}=-r^2, \label{56}\\
&g_{33}=-r^2\mbox{sin}^2\theta \label{57}
\end{align}
with 
\begin{align}
&\lambda M_1(t)=-M+\frac{4\pi}{kc^2}\frac{ (\varepsilon +2)6^\varepsilon 
(kc^2M)^{\varepsilon +1}}{(8\pi)^{(\varepsilon +1)}[R_o(t)]^{3\varepsilon}},  \label{57a}\\
&\lambda M_2(t)=\frac{4\pi}{kc^2}\frac{\varepsilon (\varepsilon +1) 
(3kc^2M)^{\varepsilon -1}\left[ \frac{\partial}{\partial t}\frac{M}{[R_o(t)]^3}\right]^2 
\alpha (t)}{(4\pi)^{\varepsilon -1}[R_o(t)]^{3\varepsilon -3}}. \label{57b}
\end{align}
We see, for example, in \eqref{50} or \eqref{54}, that the  metric in an $f(R)$-gravity is 
different from the one in Einstein's GR by last two terms. If  the body-gravitation source 
shrinks or expands (it means that its radius depends on time), the  metric would depend on 
time, unlike the Einstein equation giving no such a effect.\\ 
\textbf{\subsection{General perturbative solution}}
In the previous subsection, vacuum solutions have been found for an arbitrary and some more 
special $f(R)$, now we will look for a general solution everywhere, not only in vacuum. 
Inside matter we do not have $ u(r,t)=-v(r,t) $, thus we will solve this problem in the 
following way: Doing the same calculations for obtaining formula \eqref{22} we get
\begin{align}
\square^E h'(kT)-\nabla^1\nabla^E_1h'(kT)=\beta (r,t), \label{58}
\end{align}
where,
\begin{align}
\beta (r.t)=&\frac{g^{00}_E}{c^2}\frac{\partial^2h'(kT)}{\partial t^2}
+\frac{1}{2c^2}\frac{\partial g^{00}_E}{\partial t}\frac{\partial 
h'(kT)}{\partial t}\nonumber \\
&+g^{11}_E\left( \frac{2}{r}-\frac{1}{2g^{00}_E}\frac{\partial 
g^{00}_E}{\partial r}\right) \frac{\partial h'(kT)}{\partial r}. \label{59}
\end{align}
The index $ E $ means the  metric tensor taken within the  Einstein theory 
(see the appendix for its values inside or outside the body-gravitation source). 
Substituting \eqref{58} into \eqref{9} we obtain the equation 
\begin{align}
R^{1}_{~1}-\frac{1}{2}R=&-kT^{1}_{~1}+\frac{\lambda}{2}h(kT) +\lambda 
k(T^{1}_{~1}-\frac{1}{2}T)h'(kT)\nonumber \\
&+\lambda \beta (r,t), \label{60}
\end{align}
which with using \eqref{29} leads to
\begin{align}
e^{-v(r,t)}\left[ \frac{u'(r,t)}{r}+\frac{1}{r^2}\right] 
-\frac{1}{r^2}=-kT^{1}_{~1}+\frac{\lambda}{2}h(kT)\nonumber \\
+\lambda k(T^{1}_{~1}-\frac{1}{2}T)h'(kT)
+\lambda \beta (r,t) \label{61}
\end{align}
or
\begin{align}
u'(r,t)=&r\left\lbrace e^{v(r,t)}\left[ \frac{1}{r^2}-kT^{1}_{~1}
+\frac{\lambda}{2}h(kT)\right.\right.\nonumber \\
&\left.\left. +\lambda k(T^{1}_{~1}-\frac{1}{2}T)h'(kT)+\lambda \beta (r,t)\right] 
-\frac{1}{r^2}\right\rbrace. \label{62}
\end{align}
Since $ e^{v(r,t)}=-g_{11}(r,t) $, we can re-write the latter equation as 
\begin{align}
u'(r,t)=&rg_{11}(r,t)\left[ -\frac{1}{r^2}+kT^{1}_{~1}-\frac{\lambda}{2}h(kT) \right.\nonumber \\
&\left.-\lambda k(T^{1}_{~1}+\frac{1}{2}T)h'(kT) 
-\lambda \beta (r,t)\right] -\frac{1}{r}. \label{63}
\end{align}
Doing integration of the above equation and noticing that $g_{00}(r,t)\rightarrow 1 $ 
as $r\rightarrow \infty$, we have
\begin{align}
u(r,t)=&\int_\infty^r\left\lbrace r'g_{11}(r',t)\left[ -\frac{1}{r'^2}+kT^{1}_{~1}
-\frac{\lambda}{2}h(kT)\right.\right.\nonumber \\
&\left.\left. -\lambda k(T^{1}_{~1}+\frac{1}{2}T)h'(kT)-\lambda \beta (r',t)\right] 
-\frac{1}{r'}\right\rbrace dr'.  \label{64}
\end{align}
Thus, from \eqref{11}, \eqref{18} and \eqref{64} the metric gets the form 
\begin{align}
g_{00}(r,t)=&\exp\int_\infty^r\left\lbrace r'g_{11}(r',t)\left[ -\frac{1}{r'^2}
+kT^{1}_{~1}-\frac{\lambda}{2}h(kT) 
\right.\right.\nonumber \\
&\left.\left.-\lambda k(T^{1}_{~1}+\frac{1}{2}T)h'(kT)-\lambda \beta (r',t)\right] 
-\frac{1}{r'}\right\rbrace dr',  \label{65}
\end{align}
\begin{align}
g_{11}(r,t)=&-\left\{1-\frac{1}{r}\int_0^r\left[ kT^0_{~0}-\frac{\lambda}{2}h(kT)
\right.\right.\nonumber \\
&-\lambda k(T^0_{~0}-\frac{T}{2})h'(kT)\nonumber \\ 
&\left.\left.-\lambda \nabla^i \nabla_i^E h'(kT)\right] r'^2dr'\right\}^{-1}, \label{66}
\end{align}
\begin{align}
g_{22}=-r^2, \label{67}
\end{align}
\begin{align}
g_{33}=-r^2\sin^2\theta, \label{68}
\end{align}
where $ \nabla^i \nabla_i^E h'(kT) $ and $ \beta(r',t) $ are given by  \eqref{22} and 
\eqref{59}, respectively.\\
\section{Motion in a central field}
In this section we will apply the obtained solutions to a motion in a central field, for example, 
a planetary motion around an isotropic star (which can be a normal star, neutron star, black hole, 
or other body-gravitational sources). This central field is not necessarily static, the radius of 
the star can expand or shrink during the time. Here we only take the models which satisfy $ h(0)=0 $, 
meaning that $ h(kT^0_{~0})=0 $, in vacuum [the models II and III, considered in Subsects. 2.1.2 
and 2.1.3, respectively, satisfy  this, but the model I (in Subsect. 2.1.1) does not]. With these 
models,  the integrations in \eqref{38} -- \eqref{41} done only within the radius $ R_0(t) $ of the 
star, lead to the solution  
\begin{align}
&g_{00}(r,t)=1-\frac{kc^2[M-\lambda M_1(t)-\lambda M_2(t)]}{4\pi r}, \label{a70}\\
&g_{11}(r,t)=\frac{-1}{1-\frac{kc^2[M-\lambda M_1(t)-\lambda M_2(t)]}{4\pi r}}, \label{a71}\\
&g_{22}=-r^2, \label{a72}\\
&g_{33}=-r^2\sin^2\theta, \label{a73}
\end{align}
where
\begin{align}
&M_1(t)=\frac{2\pi [R_o(t)]^3}{3kc^2}\left[h(kT^0_{~0})+kT^0_{~0}h'(kT^0_{~0})\right],  \label{a74}\\
&M_2(t)=\frac{4\pi}{kc^2}h''(kT^0_{~0})\left[ \frac{\partial}{\partial t}
\frac{M}{[R_o(t)]^3}\right]^2 ~\alpha (t),  \label{a75}
\end{align}
with  $ T^0_{~0} $ calculated by  \eqref{24}. For the model II,  
$M_1(t)$ and $M_2(t)$ have the form 
\begin{align}
&M_1(t)=\frac{4\pi}{kc^2}\frac{(b+1)c^{2b}(kM)^b }{3^{1-b}~2^{2b+1}~{\pi}^b[R_o(t)]^{3b-3}}, \label{a76}\\
&M_2(t)=\frac{4\pi}{kc^2}\frac{b(b-1)c^{2b-4}(3kM)^{b-2}\left[ 
\frac{\partial}{\partial t}\frac{M}{[R_o(t)]^3}\right]^2 \alpha (t)}{(4\pi)^{b-2}[R_o(t)]^{3b-6}}. \label{a77}
\end{align}
and for the model III they become 
\begin{align}
&\lambda M_1(t)=-M+\frac{4\pi}{kc^2}\frac{ (\varepsilon +2)6^\varepsilon 
(kc^2M)^{\varepsilon +1}}{(8\pi)^{(\varepsilon +1)}[R_o(t)]^{3\varepsilon}},  \label{a78}\\
&\lambda M_2(t)=\frac{4\pi}{kc^2}\frac{\varepsilon (\varepsilon +1) 
(3kc^2M)^{\varepsilon -1}\left[ \frac{\partial}{\partial t}\frac{M}{[R_o(t)]^3}\right]^2 
\alpha (t)}{(4\pi)^{\varepsilon -1}[R_o(t)]^{3\varepsilon -3}}. \label{a79}
\end{align}
Setting 
\begin{align}
M_f(t)=M-\lambda M_1(t)-\lambda M_2(t) \label{a80}
\end{align}
and noticing that $ k=\frac{8\pi G}{c^4}$, 	we have a Schwarzschild-type metric 
\begin{align}
ds^2=&\left[ 1-\frac{2GM_f(t)}{c^2r}\right] {dx^0}^2-\frac{dr^2}{1-\frac{2GM_f(t)}{c^2r}}\nonumber \\
&-r^2(d\theta^2+\sin^2\theta{d\varphi^2}). \label{a81}
\end{align}
Here, $M_f$ can be treated as an effective mass in an $f(R)$-gravity, which, then, looks like 
the GR in a central field of a source with a non-static mass $M_f$. In general, it is a 
function of time 
even when the mass $M$ is a constant. This may lead to interesting phenomena 
which can be discussed elsewhere later.  
Let us make a coordinate transformation changing $r$ as 
\begin{align}
r\longrightarrow r\left[ 1+\frac{GM_f(t)}{2c^2r}\right] ^2 \label{a82},
\end{align}
but keeping other coordinates unchanged  ($ t\longrightarrow t $, $\theta\longrightarrow 
\theta  $, $ \varphi\longrightarrow \varphi $). Subsituting \eqref{a82} into \eqref{a81} 
and neglecting the infinitesimal terms  
\begin{align*}
\frac{2 G}{c^2}\frac{\partial M_f(t)}{\partial t}\frac{\left[ 1+
\frac{GM_f(t)}{2c^2r}\right] ^4}{1-\frac{GM_f(t)}{2c^2r}}dtdr,
\end{align*}
and 
\begin{align*}
\frac{G^2}{c^6}\left(\frac{\partial M_f(t)}{\partial t} \right)^2 \frac{\left[ 1
+\frac{GM_f(t)}{2c^2r}\right] ^4}{\left[ 1-\frac{GM_f(t)}{2c^2r}\right] ^2}{dx^0}^2,
\end{align*}
very small compared with  
$$\frac{\left[ 1-\frac{GM_f(t)}{2c^2r}\right] ^2}{\left[ 1+\frac{GM_f(t)}{2c^2r}\right] ^2}{dx^0}^2,$$
we get 
\begin{align}
ds^2=&\frac{\left[ 1-\frac{GM_f(t)}{2c^2r}\right] ^2}{\left[ 1+\frac{GM_f(t)}{2c^2r}\right] ^2}{dx^0}^2
-\left[ 1+\frac{GM_f(t)}{2c^2r}\right] ^4
\nonumber \\
&\times (dr^2+r^2d\theta^2+r^2sin^2\theta{d\varphi^2}). \label{a83}
\end{align}
Since  $dr^2+r^2d\theta^2+r^2sin^2\theta{d\varphi^2}=dx^2+dy^2+dz^2$, we have $ g_{xx}=g_{yy}= g_{zz}$. 
It means that the spacial coordinates $x$, $y$, $z$  play the same role in the isotropic frame.\\

From the special relativity, we know the Hamilton-Jacobi equation of a free particle in a flat space-time,
\begin{align}
g^{\mu\nu}\frac{\partial{S}}{\partial{x^\mu}}\frac{\partial{S}}{\partial{x^\nu}}=m^2c^2,\label{a84}
\end{align}
where $m$ and $S$ are its mass and action, respectively \cite{2,21a}. 
Since $S$ is a scalar, the equation \eqref{a84} is still valid for the general relativity, where the 
flat space-time is replaced by a curved space-time.
 In the metric \eqref{a83}, the equation \eqref{a84} has the form 
\begin{align}
&\left[ \frac{1+\frac{GM_f(t)}{2c^2r}}{1-\frac{GM_f(t)}{2c^2r}}\right]^2\left( \frac{\partial{S}}
{\partial{x^0}}\right)^2-\frac{1}{\left[ 1+\frac{GM_f(t)}{2c^2r}\right] ^4}\left[\left 
(\frac{\partial{S}}{\partial{r}}\right )^2\right.\nonumber \\
&\left.+\frac{1}{r^2}\left (\frac{\partial{S}}{\partial{\theta}}\right )^2
+\frac{1}{r^2sin^2\theta}\left (\frac{\partial{S}}{\partial{\varphi}}\right )^2\right] =m^2c^2. \label{a85}
\end{align}
Let us apply this equation to a planet's motion around an isotropic star producing a central gravitational 
field. Since the planet moves in a fixed plane passing the star's center taken for the origin of the 
coordinate frame, we can choose the orientation of the coordinate frame so that the planet's orbital plane 
is horizontal, that is, we always have $\theta=\frac{\pi}{2}$ and, thus,     
\begin{align}
&\left[ \frac{1+\frac{GM_f(t)}{2c^2r}}{1-\frac{GM_f(t)}{2c^2r}}\right] ^2\left (\frac{\partial{S}}
{\partial{x^0}}\right )^2-\frac{1}{\left[ 1+\frac{GM_f(t)}{2c^2r}\right] ^4}\left[\left 
(\frac{\partial{S}}{\partial{r}}\right )^2\right.\nonumber \\
&\left.+\frac{1}{r^2}\left (\frac{\partial{S}}{\partial{\varphi}}\right )^2\right] =m^2c^2.  \label{a86}
\end{align}
Because 
\begin{align}
\frac{\partial S}{\partial t}=-H(t), \label{a87}
\end{align}
where $H(t)$ is the Hamiltonian, it follows that
\begin{align}
&\left[ \frac{1+\frac{GM_f(t)}{2c^2r}}{1-\frac{GM_f(t)}{2c^2r}}\right] ^2\left[ \frac{H(t)}{c}\right] ^2
-\frac{1}{\left[ 1+\frac{GM_f(t)}{2c^2r}\right] ^4}\left[\left (\frac{\partial{S}}{\partial{r}}\right )^2\right.\nonumber \\
&\left.+\frac{1}{r^2}\left(\frac{\partial{S}}{\partial{\varphi}}\right)^2\right] =m^2c^2.  \label{a88}
\end{align}
In a central field which may not be static, the Hamiltonian may not be conserved but the angular momentum 
is always conserved. Following \cite{21a}, we write  
\begin{align}
\frac{\partial S}{\partial \varphi}=p_{\varphi}=\mu, \label{a89}
\end{align}
where $ \mu $  is the conserved value of the angular momentum $p_{\varphi}$,  
or 
\begin{align}
S=\mu \varphi +s(r,t), \label{a90}
\end{align}
where $s(r,t)$ is a function of $r$ and $t$ only. 
Taking \eqref{a90} into account, Eq. \eqref{a88} becomes
\begin{align}
\left[ \frac{1+\frac{GM_f(t)}{2c^2r}}{1-\frac{GM_f(t)}{2c^2r}}\right] ^2\left[ \frac{H(t)}{c}\right] ^2
-\frac{\left[ \frac{\partial{s(r,t)}}{\partial{r}}\right] ^2
+\frac{\mu^2}{r^2}}{\left[ 1+\frac{GM_f(t)}{2c^2r}\right] ^4}= m^2c^2.  \label{a91}
\end{align}
Solving the latter equation for $s(r,t)$ we get 
\small
\begin{align}
&s(r,t)=\nonumber\\
&\int\left\lbrace 
\left[ 1+\frac{GM_f(t)}{2c^2r}\right] ^4\left( \left[ \frac{1+\frac{GM_f(t)}{2c^2r}}
{1-\frac{GM_f(t)}{2c^2r}}\right]^2\left[ \frac{H(t)}{c}\right] ^2-m^2c^2\right) \right.\nonumber \\
&\left.-\frac{\mu^2}{r^2}\right\rbrace ^{1/2}~dr.  \label{a93}
\end{align}
\normalsize 
Note that in the above integral the coordinates $ r$ and $t $  are treated as independent variables. 
As $\frac{\partial S}{\partial \mu}$ is also a constant of motion \cite{2,21a}, 
\begin{align}
\frac{\partial S}{\partial \mu}= \mbox{const}., \label{a94}
\end{align}
it means   
\begin{align}
\varphi =-\frac{\partial s(r,t)}{\partial \mu}+\mbox{const}.. \label{a95}
\end{align}
Combining \eqref{a93} with  \eqref{a95}, we find 
\begin{align}
\varphi =\int \frac{\mu /r^2 dr}{\sqrt{\frac{ \left[ \frac{1+\frac{GM_f(t)}{2c^2r}}{1
-\frac{GM_f(t)}{2c^2r}}\right]  ^2\left[ \frac{H(t)}{c}\right] ^2-m^2c^2 }{\left[ 
1+\frac{GM_f(t)}{2c^2r}\right] ^{-4}}~~-~~\frac{\mu^2}{r^2}}}. \label{a96}
\end{align}
\textbf{\subsection{Motion of a planet in a central field of a star}}
Let us consider the motion of a planet around an isotropic star. 
If we write the Hamiltonian in the form 
\begin{align}
H(t)=E(t)+mc^2, \label{a97}
\end{align}
[where  $ E(t) $ has the meaning of both kinetic energy and potential energy of the planet in 
the gravitational field], then  \eqref{a96} is rewritten as 
\begin{align}
\varphi =\int \frac{\mu /r^2dr}{\sqrt{\frac{ \left[ \frac{1+\frac{GM_f(t)}{2c^2r}}{1
-\frac{GM_f(t)}{2c^2r}}\right]  ^2\left[ \frac{E^2(t)+2mc^2E(t)+m^2c^4}{c^2}\right]
-m^2c^2}{\left[ 1+\frac{GM_f(t)}{2c^2r}\right] ^{-4}} -\frac{\mu^2}{r^2}}}. \label{a98}
\end{align}
On the other hand, if we consider the planet's motion to be relatively slow (compared with the light), 
$\frac{v^2}{c^2}\ll 1$, where, $v$ is the planet's speed, that is, $E^2\ll |2mc^2E|$, 
we have
\begin{align}
\varphi =\int \frac{\mu /r^2dr}{\sqrt{\frac{ \left[ \frac{1+\frac{GM_f(t)}{2c^2r}}{1
-\frac{GM_f(t)}{2c^2r}}\right]  ^2\left[ 2mE(t)+m^2c^2\right]-m^2c^2}{\left[ 
1+\frac{GM_f(t)}{2c^2r}\right] ^{-4}} -\frac{\mu^2}{r^2}}},  \label{a99}
\end{align}
or
\begin{align}
&\varphi =\int dr \nonumber\\
& \frac{\mu /r^2}{\sqrt{\frac{2mE(t) \left[ \frac{1+\frac{GM_f(t)}{2c^2r}}{1
-\frac{GM_f(t)}{2c^2r}}\right] ^2}{\left[ 1+\frac{GM_f(t)}{2c^2r}\right] ^{-4}} 
+\frac{2m^2GM_f(t)}{r} \frac{ \left[ 1+\frac{GM_f(t)}{2c^2r}\right] ^4}{\left[ 
1-\frac{GM_f(t)}{2c^2r}\right] ^2}  -\frac{\mu^2}{r^2}}}.  \label{a100}
\end{align}
At the first order approximation  
\begin{align}
\left[ \frac{1+\frac{GM_f(t)}{2c^2r}}{1-\frac{GM_f(t)}{2c^2r}}\right] ^2\cong\left[ 
1+\frac{GM_f(t)}{2c^2r}\right] ^4 \cong1+\frac{2GM_f(t)}{c^2r}, \label{a101}
\end{align}
the equation \eqref{a100} takes the form 
\begin{align}
\varphi&= \int\frac{\mu/r^2dr}{\sqrt{\frac{2mE(t)}{\left[ 1+\frac{4GM_f(t)}{c^2r}\right]^{-1}} 
+\frac{\frac{2m^2GM_f(t)}{r}}{\left[ 1+\frac{3GM_f(t)}{c^2r}\right]^{-1}} -\frac{\mu^2}{r^2}}},  \label{a102}
\end{align}
that is, 
\begin{align}
&\varphi= \int dr\nonumber\\
&\frac{\mu/r^2}{\sqrt{\frac{2m}{\left\lbrace E(t)+\frac{mGM_f(t)}{r}\left[ 
1+\frac{4E(t)}{mc^2}\right] \right\rbrace^{-1}} -\frac{\frac{\mu^2}{r^2}}{\left[ 
1-\frac{6m^2G^2M_f^2(t)}{c^2\mu^2}\right]^{-1}} }}. \label{a103}
\end{align}
Using the notation
\begin{align}
\beta(t)= ~& mGM_f(t)\left[ 1+\frac{4E(t)}{mc^2}\right],  \label{a104}\\
l^2(t)= ~& \mu^2\left[ 1-\frac{6m^2G^2M_f^2(t)}{c^2\mu^2}\right],  \label{a105}
\end{align}
we obtain the formula  
\begin{align}
\varphi=\frac{1}{\sqrt{1-\frac{6m^2G^2M_f^2(t)}{c^2\mu^2}}}\int\frac{l(t)/r^2}{\sqrt{2m\left[ 
E(t)+\frac{\beta(t)}{r}\right] -\frac{l^2(t)}{r^2}}}dr, \label{a106}
\end{align}
which after doing integration becomes  
\begin{align}
\varphi=\frac{1}{\sqrt{1-\frac{6m^2G^2M_f^2(t)}{c^2\mu^2}}}\mbox{arccos}\frac{\frac{l(t)}{r}
-\frac{m\beta(t)}{l(t)}}{\sqrt{2mE(t)+\frac{m^2\beta^2(t)}{l^2(t)}}}+ C, 
\end{align}
where $C$ is a constant. 
Rotating the coordinate frame 
so that $C = 0$,   
\begin{align}
\varphi=\frac{1}{\sqrt{1-\frac{6m^2G^2M_f^2(t)}{c^2\mu^2}}}\mbox{arccos}\frac{\frac{l(t)}{r}
-\frac{m\beta(t)}{l(t)}}{\sqrt{2mE(t)+\frac{m^2\beta^2(t)}{l^2(t)}}}, \label{a108}
\end{align}
we get  
\begin{align}
&\frac{l^2(t)}{m\beta(t)r}=\nonumber\\
&1+\sqrt{1+\frac{2E(t)l^2(t)}{m\beta^2(t)}}\mbox{cos}\left( \sqrt{1-\frac{6m^2G^2M_f^2(t)}{c^2\mu^2}}\varphi\right). \label{a109}
\end{align}
This is the equation of motion of a planet in a central field of a star. We notice that $t=0$ is chosen arbitrarily  
after having ``gravitational interaction" passing the planet. The equation \eqref{a106} or \eqref{a108} or  
\eqref{a109} is a general equation of  a planetary motion in a central field of a star.\\

\begin{figure}[h]
\begin{center}
\includegraphics[scale=0.2]{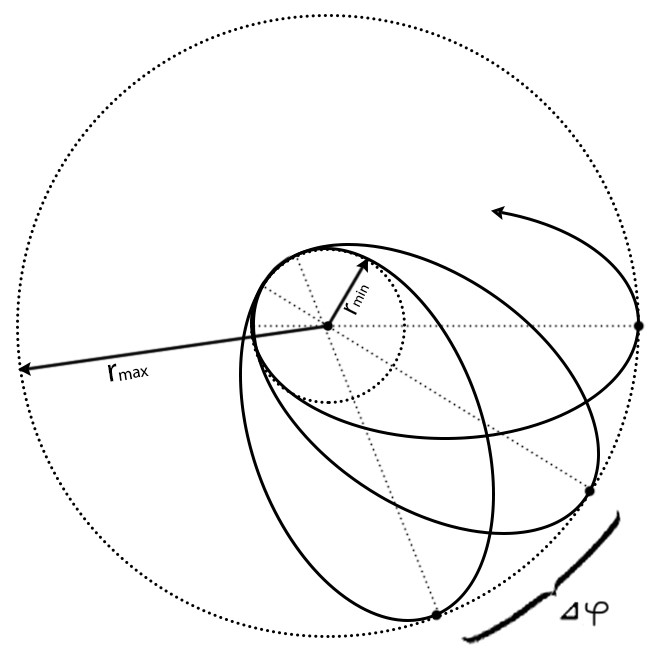}
\caption{\label{fig:1}\textit{In a static central field, both  $r_e$ and  $ \Delta \varphi $ remain  
unchanged over time as in Einstein's theory but have a correction to the corresponding Einstein's 
values (illustration following \cite{21a})}.}
\end{center}
\end{figure}

Now we consider a planet moving in a nearly-elliptic orbit. From \eqref{a109} we can see that the lengths of the major 
axis and the minor axis of the near-elliptic orbit change if the central filed is not static (note that the central 
field is not static even when the total mass $M$ of the star is unchanged, if the star expands or shrinks keeping its 
isotropic form). This is an effect which cannot occur in the Einstein theory [when the radius of the star expands or 
shrinks, the metric in the vacuum in the Einstein theory does not depend on time]. The extremums (apsides) $r_{e}$ of 
$r$, which are its minimal value (periastron/perihelion) $r_{p}$ or maximal value (apastron/aphelion) $r_{a}$,  
can be  calculated as follows: First, we notice that the argument of an arccos fuction varies within the interval [-1,1], 
hence from \eqref{a108} we have the condition 
\begin{align}
-1 \leqslant \frac{\frac{l(t)}{r}-\frac{m\beta(t)}{l(t)}}{\sqrt{2mE(t)+\frac{m^2\beta^2(t)}{l^2(t)}}} \leqslant 1.  
\label{a110}
\end{align} 
The extremums $r_{e}$ are found at the two edges of this interval,
\begin{align}
\frac{\frac{l(t_{e})}{r_{e}}-\frac{m\beta(t_{e})}{l(t_{e})}}{\sqrt{2mE(t_{e})+\frac{m^2\beta^2(t_{e})}{l^2(t_{e})}}}=\pm 1,  
\label{a111}
\end{align}
where, $t_e$ is the time value corresponding to $r_e$. Hence,  
\begin{align}
r_{e}=&\frac{l^2(t_{e})}{m\beta(t_{e})\pm\sqrt{m^2\beta^2(t_{e})+2mE(t_{e})l^2(t_{e})}} \label{re}
\end{align}
or, more precisely, 
\begin{align}
r_{p}=&\frac{l^2(t_{p})}{m\beta(t_{p})+\sqrt{m^2\beta^2(t_{p})+2mE(t_{n})l^2(t_{p})}}, \label{a112}\\
r_{a}=&\frac{l^2(t_{a})}{m\beta(t_{a})-\sqrt{m^2\beta^2(t_{a})+2mE(t_{a})l^2(t_{a})}}. \label{a113}
\end{align}
From \eqref{a108}, we have
\begin{align}
\varphi_e=\frac{1}{\sqrt{1-\frac{6m^2G^2M_f^2(t_e)}{c^2\mu^2}}}\mbox{arccos}\frac{\frac{l(t_e)}{r_e}
-\frac{m\beta(t_e)}{l(t_e)}}{\sqrt{2mE(t_e)+\frac{m^2\beta^2(t_e)}{l^2(t_e)}}}, \label{a114}
\end{align}
and with \eqref{re} substituted into \eqref{a114} we obtain
\begin{align}
\varphi_e=\frac{1}{\sqrt{1-\frac{6m^2G^2M_f^2(t_e)}{c^2\mu^2}}}\mbox{arccos}(\pm 1), \label{a115}
\end{align}
for $e$ being $p$ or $a$ (but not both simultaneously). 
It follows that
\begin{align}
\varphi_e(k)=\frac{k\pi}{\sqrt{1-\frac{6m^2G^2M_f^2(t_k)}{c^2\mu^2}}}~~~~~(k\in \ZZ), \label{a116}
\end{align}
with  $\varphi_e(k)$ being the set of all the values of the angle $\varphi$ at which  $r$  gets an extremum value $r_e$, 
where, $k$ is even for $e=p$ and odd for $e=a$. Thus, the orbital precession is  
\begin{align}
&\Delta \varphi_e (k) = \varphi_e (k+1)- \varphi_e (k)=\nonumber\\
&\frac{2(k+1)\pi}{\sqrt{1-\frac{6m^2G^2M_f^2(t_{k+1})}{c^2\mu^2}}}-\frac{2k\pi}{\sqrt{1
-\frac{6m^2G^2M_f^2(t_k)}{c^2\mu^2}}} ~ (\text{mod} ~ 2\pi).\label{a117}
\end{align}
If $ M_f(t_{k+1})\cong M_f(t_{k})$, 
Eq. \eqref{a117} can be taken approximately as 
\begin{align}
\Delta \varphi_e (k) &\cong \frac{2\pi}{\sqrt{1-\frac{6m^2G^2M_f^2(t_k)}{c^2\mu^2}}} ~ (\text{mod} ~ 2\pi)
\nonumber\\
&\cong 2\pi\left[ 1+\frac{3m^2G^2M_f^2(t_k)}{c^2\mu^2}\right] ~ (\text{mod} ~ 2\pi). \label{a119}
\end{align}
Thus, the orbital precession becomes 
\begin{align}
\Delta \varphi_e (k)=\frac{6\pi m^2G^2M_f^2(t_k)}{c^2\mu^2}.\label{Detaphi}
\end{align} 
Therefore, the correction to Einstein's precession  is
\begin{align}
&\delta \varphi_e (k) =\nonumber\\
&\frac{6\pi m^2G^2\{\lambda^2[M_1(t_k)+M_2(t_k)]^2-2\lambda M[M_1(t_k)+M_2(t_k)]\}}{c^2\mu^2},
\label{deltaphi}
\end{align}
or
\begin{align}
\delta \varphi_e (k) \cong 
\frac{-12\pi m^2G^2\lambda M[M_1(t_k)+M_2(t_k)]}{c^2\mu^2}
\label{deltaphi2}
\end{align}
when the second order approximation is neglected. 
If the central field is static, $\Delta \varphi_e$ (thus, $\delta \varphi_e$) is a constant 
(see Figure \ref{fig:1} which is only illustrative), 
but when the central field is not static (for example when the radius of the star expands or shrinks 
keeping the star's isotropy) $\Delta \varphi_e$ (thus, $\delta \varphi_e$) is no longer a constant, 
and there will be new effects compared with Eisntein's theory:
\begin{figure}[h]
\begin{center}
\includegraphics[scale=0.2]{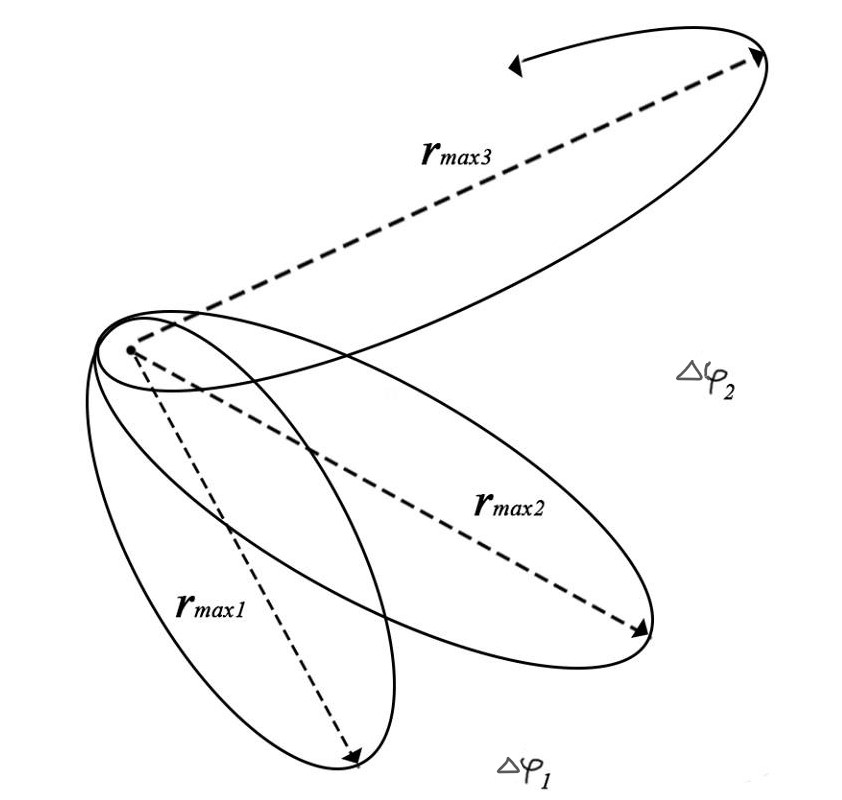}
\caption{\label{fig:2}\textit{In a non-static case, both $r_e$ and  $ \Delta \varphi $ change 
over time, unlike the corresponding Einstein's values remaning always unchanged in a central 
field (from a source with a constant mass)}.}
\end{center}
\end{figure}
there is not only a corrected orbital precession given by \eqref{Detaphi} and \eqref{deltaphi}, 
but, as shown in \eqref{a112} and \eqref{a113}, the axes of the orbit also varies with time 
(see an illustration in Figure \ref{fig:2}). 

\textbf{\subsection{Light's propagation in the central field of a star}}
Since $ m=0 $ in this case, Eq. \eqref{a96} is reduced to 
\begin{align}
\varphi =\int \frac{\mu /r^2}{\sqrt{\left[ 1+\frac{GM_f(t)}{2c^2r}\right] ^4 \left[ 
\frac{1+\frac{GM_f(t)}{2c^2r}}{1-\frac{GM_f(t)}{2c^2r}}\right]  ^2\left[ 
\frac{H(t)}{c}\right] ^2 -\frac{\mu^2}{r^2}}}dr. \label{a121}
\end{align}
Then, using \eqref{a101}, we obtain
\begin{align}
\varphi =\int \frac{\mu /r^2}{\sqrt{\left[ 1+\frac{4GM_f(t)}{c^2r}\right] \left[ \frac{H(t)}{c}\right] 
^2 -\frac{\mu^2}{r^2}}}dr. \label{a122}
\end{align}
At the first order of approximation, it follows that
\begin{align}
\varphi =&\int \frac{\mu /r^2}{\sqrt{\frac{H^2(t)}{c^2}-\left[\frac{2GM_f(t)H^2(t)}{c^4\mu}  
-\frac{\mu}{r}\right] ^2}}dr, \label{a123}
\end{align}
or
\begin{align}
\varphi = \mbox{arccos}\left[ \frac{\mu c}{H(t) r}-\frac{2GM_f(t)H(t)}{c^3\mu}\right] + C',\label{a124}
\end{align}
where $C'$ is a constant. 
If the reference frame is chosen so that $\varphi=0$  at the point (periastron/perihelion) 
where the light is closest to the origin of the frame  
(the gravitational source), we can take $C' = 0$. 
Thus, the equation of motion of light in a central  field is 
\begin{align}
\varphi = \mbox{arccos}\left[ \frac{\mu c}{H(t) r}-\frac{2GM_f(t)H(t)}{c^3\mu}\right], \label{a125}
\end{align}
or
\begin{align}
\mbox{cos}\varphi= \frac{\mu c}{H(t) r}-\frac{2GM_f(t)H(t)}{c^3\mu}.  \label{a126}
\end{align}
At a far-distance (with $ r \equiv r_{l}$ very large), this equation has the approximate form  
\begin{align}
\mbox{cos}\varphi_{l}=-\frac{2GM_f(t_{l})H(t_{l})}{c^3\mu}.  \label{a127}
\end{align}
It means that $ \varphi_{l} > \frac{\pi}{2} $, that is, the light's trajectory is deflected toward 
the gravitational source, and the deflection angle, denoted by $\phi$, is  
\begin{align}
\phi =&2\left( \varphi_{l} - \frac{\pi}{2}\right) \label{a128}\\
=&2 \varphi_{l}-\pi. \label{a129}
\end{align}
This angle \eqref{a129} can be calculated approximately as follows. 
Denoting the periastron distance 
by $ D $ (a notation used in \cite{2}). 
As $ \varphi =0 $ at $ r=D $, from \eqref{a126} we have
\begin{align}
\frac{\mu c}{H(t_D) D}-\frac{2GM_f(t_D)H(t_D)}{c^3\mu}=1,  \label{a130}
\end{align}
it follows that
\begin{align}
\mu =\frac{H(t_D) D}{2c}\left(1+\sqrt{1+\frac{8GM_f(t_D)}{c^2D}} \right). \label{a131}
\end{align}
Neglecting again higher orders of approximation, we have
 \begin{align}
\mu 
\cong &\frac{H(t_D) D}{c}\left( 1+\frac{2GM_f(t_D)}{c^2D} \right), \label{a133}
\end{align}
or 
\begin{align}
\frac{1}{\mu}\cong \frac{c}{H(t_D) D}\left(  1-\frac{2GM_f(t_D)}{c^2D}\right).  \label{a134}
\end{align}
 Substituting \eqref{a134} into \eqref{a127} we obtain
 \begin{align}
 \mbox{cos}\varphi_{l}=-\frac{2GM_f(t_l)}{c^2D}\frac{H(t_l)}{H(t_D)}
 +\frac{4G^2M_f(t_a)M_f(t_D)}{c^4D^2}\frac{H(t_l)}{H(t_D)}, \label{a135}
 \end{align}
 where, the second term is very small compared with the first term. Thus, we can take
 \begin{align}
 \mbox{cos}\varphi_{l}=-\frac{2GM_f(t_l)}{c^2D}\frac{H(t_l)}{H(t_D)}. \label{a136}
 \end{align}
 From  \eqref{a136} we see the RHS term is very small and $ \varphi_l  > \frac{\pi}{2} $, and we can set
 \begin{align}
 \varphi_l = \frac{\pi}{2}+\varepsilon   \label{a137}
 \end{align}
 with  $ \varepsilon $ being very small. Combining \eqref{a136} with \eqref{a137}, we have 
 \begin{align}
 \cos\left( \frac{\pi}{2}+\varepsilon \right)=-\mbox{sin}\varepsilon \cong 
 -\frac{2GM_f(t_l)}{c^2D}\frac{H(t_l)}{H(t_D)}, \label{a138}
 \end{align}
 or
  \begin{align}
\varepsilon \cong \frac{2GM_f(t_l)}{c^2D}\frac{H(t_l)}{H(t_D)}, \label{a139}
 \end{align}
 and, thus,  
  \begin{align}
 \varphi_l = \frac{\pi}{2}+ \frac{2GM_f(t_l)}{c^2D}\frac{H(t_l)}{H(t_D)}.  \label{a140}
 \end{align}
 From here, taking \eqref{a129} into account, we have 
 \begin{align}
 \phi = \frac{4GM_f(t_l)}{c^2D}\frac{H(t_l)}{H(t_D)}.  \label{a141}
 \end{align}
 We can see that if the gravitational field is static  $ H(t_l)=H(t_D)=H $ then 
 $ \phi = \frac{4GM_f}{c^2D}  $ with $ M_f=M-\lambda M_1 $ [see \eqref{a80}], 
 and this angle has a correction to Einstein's value  
 \begin{align}
 \delta\phi = -\lambda \frac{4G M_1}{c^2D}.  \label{a141}
 \end{align}
In an $f(R)$ theory, the light deflection angle, like the orbital precession discussed above, 
in a static central gravitational field, has a constant correction to that in Einstein's theory, 
but when the gravitational field is not static the correction may not be a constant, but, in 
general, it depends on time. \\

\section{Conclusions}

Einstein's general theory of relativity is a triumphant theory but, as mentioned above, 
it meets several open problems such as the accelerated expansion of the Universe (or 
dark energy), the cosmic inflation, an integration with quantum theory (quantum 
gravity), etc. The $f(R)$-theory of gravity was introduced to solve some of these 
problems. Then, the Einstein equation \eqref{2} is replaced by a, more complicated in 
general, equation \eqref{6}. Usually, solving the latter is problematic and it must be 
done via an approximation method by imposing appropriate condition(s). As, physically, 
the $f(R)$-gravity is assumed to be a perturbative theory around Einstein's GR describing 
very well most of today observations, we have followed a perturbation approach to solving 
Eq. \eqref{6}. However, even with this assumption, it is not always easy to solve 
Eq. \eqref{6} without imposing any further condition. One of the most often imposed 
conditions is the spherical symmetry being a good approximation in many cases. Therefore, 
in this article we try to perturbatively solve Eq. \eqref{6} in a central field. The 
corresponding general solution 
is given in \eqref{65} -- \eqref{68}, while the vacuum solution is given in \eqref{31} -- 
\eqref{34}. At a large distance from the gravitational source the solution \eqref{31} -- 
\eqref{34} can be written in the form \eqref{38} -- \eqref{41} with some particular cases also 
considered (see 2.1.1 -- 2.1.3). These results, as discussed in Sect. 3, can be applied to 
investigating planetary and light's motions in a central field. In comparison with Einstein's 
theory, an orbital precession or a trajectory deflection now gets a correction which is a constant 
for a static central field and varies with time for a non-static central field even from a source 
of a constant mass, unlike the corresponding Einstein's value which does not change in the same 
circumstance. In general, a spherically symmetric vacuum solution of Eq. \eqref{6} is not stationary, 
while a spherically symmetric vacuum solution of the Einstein equation is always stationary. In other 
words, Birkhoff's theorem in the GR is not valid any more for a general $f(R)$-theory of gravity. 
This may have interesting consequences (for example, a spherically symmetric pulsating (or expanding 
or collapsing) object is not disabled to emit gravitational waves as in the GR) being a subject 
of our next investigation. Following the present method, we will also investigate cosmological 
equations and models corresponding to the $f(R)$-theory of gravity. \\

The results obtained above may give an indication for an experimental test of an $f(R)$-theory 
of gravity. This theory in the considered circumstance can be treated as an Einstein's GR with 
an effective mass ($M_f$) which may vary with time even in the case keeping the original total 
mass ($M$) constant. Let us make some estimation using a real data. \\

As seen above, a perturbative $f(R)$-theory can be considered as Einstein theory with an effective 
mass 
$M_f=M-\lambda M_1-\lambda M_2$ replacing the original mass $M$ assumed here to be a constant. 
According to \eqref{a75} $ M_2=0$ for a static field, this effective mass becomes 
$ M_f=M-\lambda M_1 $. Thus, from \eqref{Detaphi} we have 
$\Delta \varphi= \frac{6\pi G^2 m^2 (M-\lambda M_1)^2}{c^2 \mu^2}$ or 
\begin{equation}
\lambda M_1=M-\sqrt{\frac{c^2\mu^2}{6\pi G^2m^2}\Delta\varphi}. \label{H1}
\end{equation}
Putting $\frac{\mu^2}{GMm^2}=a(1-e^2)$ in the latter equation \eqref{H1}, where, $a$ is the length 
of a semi-major 
axis and $e$ is the eccentricity of an orbital ellipse \cite{2}, we get    
\begin{equation}
\lambda M_1=M-\sqrt{\frac{c^2M}{6\pi G}a(1-e^2)\Delta\varphi}. \label{H12}
\end{equation}
Using a recent 
data for the Mercury's orbital precession \cite{Majumder:2011eh}: 
$ c=299792458 m/s$; $G=6.67259\times 10^{-11} kg^{-1}m^3s^{-2}$; $\frac{2GM}{c^2}=2.95325008\times 10^3 m$;
$a=5.7909175\times10^{10}m$; $e=0.20563069;~\Delta\varphi_{obs}=2\pi(7.98734\pm 0,00037)\times 10^{-8}~radian/revolution $, 
we can estimate the correction $\lambda M_1 $ to the mass $M$ to be 
\begin{equation}
\lambda M_1=11.866507\times 10^{24} kg. \label{H3}
\end{equation}
It means that the Sun's mass  $ M=1.988919\times 10^{30}kg $ is effectively reduced by  
\begin{equation}
\frac{\lambda M_1}{M}=5.966309\times 10^{-6}=0.0005966309 ~\% . \label{H4}
\end{equation} 
It is quite small compared with the Sun's mass but the effect may be measurable (see below).
All these results on $\lambda M_1$ are model-independent, i.e., for 
an arbitrary $f(R)$. To estimate $\lambda$ we need, however, a concrete $f(R)$. \\ 

According to a perturbation criterion $ \lambda h(R)\ll R $ applied to \eqref{5} and taking \eqref{R} 
into account we have   
\begin{equation}
\lambda h(T)\ll \frac{8\pi G}{c^4}T, \label{H5}
\end{equation}
or for $T\approx T^0_{~0}$,
\begin{equation}
\lambda h(T^0_{~0})\ll \frac{8\pi G}{c^4}T^0_{~0}. \label{H6}
\end{equation}
Inserting $ T^0_{~0}=\frac{Mc^2}{\frac{4}{3}\pi [R_0]^3} $ (where $ R_0 $ and $ M $ are the radius and the 
mass of the body-gravitational source, respectively) we get 
\begin{equation}
\lambda h\ll \frac{6GM}{c^2[R_0]^3}. \label{H7}
\end{equation}
That means $\lambda h$ is very small, 
\begin{equation}
\lambda h\ll 26 \times 10^{-24}, \label{H7b}
\end{equation}
as expected, where the radius $R_0=6.957 \times 10^8 m $ of the Sun (see wikipedia.org) is used. 
Note that the compatibility between \eqref{H4} and \eqref{H7} depends on the model chosen. 
For example, we choose the model 
\begin{equation}
f(R)=R+\lambda R^b, \label{H7a}
\end{equation}
for $b=2$, that is, $  h(R)=R^2 $, and obeying \eqref{H7} $\lambda$ must satisfy the condition 
\begin{equation}
\lambda\ll\frac{c^2[R_0]^3}{6GM}. \label{H8}
\end{equation}
For the data given above, the latter inequality becomes  
\begin{equation}
\lambda\ll 0.380053\times 10^{23}. \label{H9}
\end{equation}
From here, we can see also $\lambda h'(R) =2\lambda R\ll 1$.
Using \eqref{H3} with $ M_1=78.4989635\times 10^6 $ calculated by \eqref{a76} for the model \eqref{H7a} 
with $b=2$ we get the following value of $\lambda$ 
\begin{equation}
\lambda=0.1511677\times 10^{18}, \label{H10}
\end{equation}
which is compatible with 
\eqref{H9}, and, thus, consistent with the observed data. From here we have   
\begin{align}
\delta\varphi= & ~\Delta \varphi_{f(R)}-\Delta \varphi_{GR}
\nonumber \\
= & -0.1906\pi \times 10^{-11} radian/revolution.
\label{phiSun}
\end{align}

As $\lambda$ may not be always very small 
the choice of $h(R)$ to satisfiy the perturbation condition is very important. For the model \eqref{H7a} 
the smaller $b>0$ is chosen the smaller $\lambda$ is obtained. For example, if $b$ has a value of the order $10^{-11}$ 
the value of $\lambda$ would be at the order $10^{-29}$. One should note that $\lambda$ is not an observable quantity 
but it can be fixed, as in \eqref{H10} for example, for a given model by using an observed data.
\\

Above, we have applied our results to the case of a motion around a star such as our Sun for which there is a very 
good experimental/observed data for reference. In order to improve the potential of an experimental detection of 
an $f(R)$-gravity effect, we can consider stronger gravitational systems like that of Sgr A* at the center of our 
Galaxy and orbiting it stars. The role of the Mercury is played now by S2 going around the "sun" Sgr A* which has 
a mass of $M=4.31\times 10^6 M_\odot=8.57\times 10^{36} kg $ and a radius of $ R_0=22\times 10^{9}m$. With the latter 
data and the orbital information of S2 ($a=0.123$ $arcsec =14.7\times 10^{13}m$ and $e=0.88$) \cite{Gillessen:2008qv}, 
we can find the deviation between the S2's orbital precessions calulated by the GR and the $f(R)$-theory as 
\begin{align}
\delta\varphi^{S2}= &~\Delta\varphi^{S2}_{f(R)}-\Delta^{S2}_{GR}\nonumber\\
= & -0.94\pi\times 10^{-6} \ radian/revolution,
\label{phiSgrA}
\end{align}
with $$\Delta\varphi^{S2}_{GR}=1.15114\pi\times 10^{-3}radian/revolution$$ calculated by the GR, and 
$$\Delta\varphi^{S2}_{f(R)}=1.1502\pi\times 10^{-3}radian/revolution$$ calculated for the model 
$f(R)=R+\lambda R^2$ by using $\lambda$ obtained in \eqref{H10} from the Sun--Mercury system. 
This deviation is much bigger than that given in \eqref{phiSun} and also bigger than the observed orbital 
precession $\Delta\varphi_{obs}$ of the Mercury, thus, much easier to be measured (with the condition that difficulties 
of measurement by other reasons, if any, are excluded or resolved).  \\

In general, the deviation between the two theories, the GR and the $f(R)$-gravity, is very small but it is measurable if one
can invent a measurement technique sensitive as that of the LIGO which is sensitive to a relative length change of an
order of around $10^{-20}$.
\section*{Acknowledgement}
This research is funded by the National Foundation for Science and Technology Development (NAFOSTED) of 
Vietnam under contract $N{\textsuperscript{\underline{o}}}$ 103.01-2017.76.
\appendix
\section
{Einstein-Schwarzchild metric inside a body-gravitational source}

Now we prove formula \eqref{36}. 
From \eqref{65} and \eqref{66} in the Einstein limit (taking $ \lambda h$ to be zero) 
we have
\begin{align}
g_{11}(r,t)=\frac{-1}{1-\frac{1}{r}\int_0^r kT^0_{~0}(r', t) r'^2dr'}, \label{Aa1}
\end{align}
\begin{align}
&g_{00}(r,t)=\nonumber\\
&\mbox{exp} \int_\infty^r\left\lbrace r'g_{11}(r',t)\left[ -\frac{1}{r'^2}+kT^{1}_{~1}(r', t)\right] 
-\frac{1}{r'}\right\rbrace dr.'  \label{Aa2}
\end{align}
Since $T^{\mu}_{~\nu}=0$ outside the gravitational source [of radius $ R_o(t) $] the latter integral becomes 
\begin{align}
&g_{00}(r,t)=\nonumber\\
&\exp \int_{R_o(t)}^r\left\lbrace r'g_{11}(r',t)
\left[ -\frac{1}{r'^2}+kT^{1}_{~1}(r', t)\right] -\frac{1}{r'}\right\rbrace dr'\nonumber \\
&\times \exp\int^{R_o(t)}_\infty \frac{ -g_{11}(r', t)-1}{r'} dr.'  \label{Aa2a}
\end{align}
On the other hand, outside the object  $ g_{11}=\frac{-1}{1-\frac{2GM}{c^2 r}} $, hence, 
\begin{align}
&g_{00}(r,t)=\nonumber\\
&\exp \int_{R_o(t)}^r\left\lbrace r'g_{11}(r',t)\left[ -\frac{1}{r'^2}+kT^{1}_{~1}(r', t)\right] 
-\frac{1}{r'}\right\rbrace dr'
\nonumber \\
&\times \left[ 1-\frac{2GM}{c^2R_o(t)}\right]. \label{Aa2c}
\end{align}
From \eqref{Aa1} and  \eqref{Aa2c}, it follows 
\begin{align}
&g_{00}(r,t)=\left[ 1-\frac{2GM}{c^2R_o(t)}\right]
\nonumber \\
&\times \exp\int_{R_o(t)}^r 
\frac{\frac{k}{r'^2}\int_0^{r'} r''^2T^{0}_{~0}(r'',t)dr'' -kr'T^{1}_{~1}(r', t)}
{1-\frac{k}{r'}\int_0^{r'} r''^2T^{0}_{~0}(r'',t)dr''}dr', \label{Aa3}
\end{align}
but, as in  \eqref{35} we consider  $ T^{0}_{~0} $ depending  on time $ t $ only 
(the density is uniform as the body-gravitation source is considered homogeneous), therefore, 
\begin{align}
&g_{00}(r,t)=\left[ 1-\frac{2GM}{c^2R_o(t)}\right]\nonumber\\
&\times\exp\int_{R_o(t)}^r 
\frac{\frac{k}{3}r'T^{0}_{~0}(t) -kr'T^{1}_{~1}(r', t)}{1-\frac{k}{3} r'^2T^{0}_{~0}(t)}dr'. \label{Aa4}
\end{align}
Treating  $ T^{1}_{~1} $ very small compared with $ T^{0}_{~0} $, we obtain
\begin{align}
g_{00}(r,t)=\left[ 1-\frac{2GM}{c^2R_o(t)}\right]\exp\int_{R_o(t)}^r 
\frac{\frac{k}{3}r'T^{0}_{~0}(t)}{1-\frac{k}{3} r'^2T^{0}_{~0}(t)}dr', \label{Aa5}
\end{align}
hence, 
\begin{align}
g_{00}(r,t)=\left[ 1-\frac{2GM}{c^2R_o(t)}\right]\sqrt{\frac{1-
\frac{k}{3}[R_o(t)]^2T^{0}_{~0}(t)}{1-\frac{k}{3}r^2T^{0}_{~0}(t)}}. \label{Aa6}
\end{align}
If   $ T^{0}_{~0} $ is considered uniform, then \eqref{Aa1} is simply 
\begin{align}
g_{11}(r,t)=\frac{-1}{1-\frac{k}{3}r^2T^{0}_{~0}(t)}. \label{Aa7}
\end{align}
Substituting  \eqref{24} into  \eqref{Aa6} and \eqref{Aa7}, we get 
\begin{align}
g_{00}(r,t)=\left[ 1-\frac{kMc^2}{4\pi R_o(t)}\right]
\sqrt{\frac{1-\frac{kMc^2}{4\pi R_o(t)}}{1-\frac{kMc^2r^2}{4\pi [R_o(t)]^3}}}, \label{Aa8}
\end{align}
\begin{align}
g_{11}(r,t)=\frac{-1}{1-\frac{kMc^2r^2}{4\pi [R_o(t)]^3}}, \label{Aa9}
\end{align}
and substituting  \eqref{Aa8} and  \eqref{Aa9} into  \eqref{35}, we get 
\begin{align}
\nabla^i \nabla_i^E h'(kT^0_{~0})\cong &\frac{k}{8\pi }
\frac{1}{\left( 1-\frac{kMc^2}{4\pi R_o(t)}\right)^{3/2}}
\frac{\partial h'(kT^0_{~0})}{\partial t} \nonumber \\
&\times \frac{r^2}{\sqrt{1-\frac{kMc^2r^2}{4\pi [R_o(t)]^3}}}
\frac{\partial}{\partial t}\ \frac{M}{[R_o(t)]^3} \nonumber\\
=& \frac{k}{8\pi}\frac{1}{\left( 1-\frac{kMc^2}{4\pi R_o(t)}\right)^{3/2}}
\frac{\partial (kT^0_{~0})}{\partial t}h''(kT^0_{~0})\nonumber \\
&\times\frac{r^2}{\sqrt{1-\frac{kMc^2r^2}{4\pi [R_o(t)]^3}}}
\frac{\partial}{\partial t} \frac{M}{[R_o(t)]^3}. \label{Aa11}
\end{align}
Then, from  \eqref{24} and  \eqref{Aa11}, it follows  
\begin{align}
\nabla^i \nabla_i^E h'(kT^0_{~0})\cong& \frac{3k^2c^2}{32\pi^2}
\frac{h''(kT^0_{~0})}{\left( 1-\frac{kMc^2}{4\pi R_o(t)}\right)^{3/2}}\nonumber \\
&\times\frac{r^2}{\sqrt{1-\frac{kMc^2r^2}{4\pi [R_o(t)]^3}}}
\left[ \frac{\partial}{\partial t}\frac{M}{[R_o(t)]^3}\right]^2, \label{Aa12}
\end{align}
therefore, 
\begin{align}
\int_o^{R_o(t)}&\nabla^i \nabla_i^E h'(kT^0_{~0})r^2dr\cong \frac{3k^2c^2}{32\pi^2}
\left[ \frac{\partial}{\partial t}\frac{M}{[R_o(t)]^3}\right]^2 \nonumber \\
&\times \frac{h''(kT^0_{~0})}{\left( 1-\frac{kMc^2}{4\pi R_o(t)}\right)^{3/2}}
\int_o^{R_o(t)}\frac{r^4}{\sqrt{1-\frac{kMc^2r^2}{4\pi [R_o(t)]^3}}}dr. \label{Aa13}
\end{align}
Denoting 
\begin{align}
\xi^2 (t)=\frac{kMc^2}{4\pi [R_o(t)]^3} \label{Aa14}
\end{align}
we re-write \eqref{Aa13} as 
\begin{align}
\int_o^{R_o(t)}&\nabla^i \nabla_i^E h'(kT^0_{~0})r^2dr\cong \frac{3k^2c^2}{32\pi^2}
\left[ \frac{\partial}{\partial t}\frac{M}{[R_o(t)]^3}\right]^2 \nonumber \\ 
&\times \frac{h''(kT^0_{~0})}{\left\lbrace 1-[\xi (t)R_o(t)]^2\right\rbrace ^{3/2}}
\int_o^{R_o(t)}\frac{r^4}{\sqrt{1-\xi^2 (t)r^2}}dr. \label{Aa15}
\end{align}
Finally, it is easy to see that 
\begin{align}
&\frac{r^4}{\sqrt{1-\xi^2r^2}}=\nonumber\\
&\frac{1}{8\xi^5}\frac{\partial}{\partial r}
\left[3\mbox{arcsin}(\xi r)-\xi r(3+2\xi^2 r^2)\sqrt{1-\xi^2r^2} \right], \label{Aa16}
\end{align}
thus, 
\small
\begin{align}
\int_o^{R_o(t)}\nabla^i \nabla_i^E h'(kT^0_{~0})r^2dr\cong h''(kT^0_{~0})
\left[ \frac{\partial}{\partial t}\frac{M}{[R_o(t)]^3}\right]^2 \alpha (t), \label{Aa17}
\end{align}
\normalsize
with 
\begin{align}
\alpha (t)=&\frac{3k^2c^2R_0(t)}{256\pi^2[\xi (t)]^4}\left\lbrace \frac{3}{\xi(t)R_0(t)}
\arcsin[\xi (t) R_0(t)]\right.\nonumber \\
 &\left. -\left( 3+2[\xi(t)R_0(t)]^2\right) \sqrt{1-[\xi(t)R_0(t)]^2}\right\rbrace
 \nonumber \\
&\times \left( 1-[\xi (t)R_0(t)]^2\right)^{-3/2}. \label{Aa18}
\end{align}

\end{document}